\documentclass[journal]{IEEEtran}

\ifCLASSINFOpdf
\else
   \usepackage [dvips]{graphicx}
\fi
\usepackage{url}

\usepackage{graphicx}
\usepackage[caption=false,font=footnotesize]{subfig}
\usepackage{amssymb}
\usepackage{amsmath}
\usepackage{xcolor}
\usepackage{multirow}
\newcommand{\rev}[1]{#1}

\ifdefined\ShowROneChanges
\newcommand{\rOne}[1]{\textcolor{red}{#1}}
\makeatletter
\let\ROneOriginalBibitem\@bibitem
\def\ROneMarkedReference{itut_p8622_2007}
\renewcommand{\@bibitem}[1]{%
  \ROneOriginalBibitem{#1}%
  \def\ROneCurrentReference{#1}%
  \ifx\ROneCurrentReference\ROneMarkedReference
    \expandafter\rOne
  \fi
}
\makeatother
\else
\newcommand{\rOne}[1]{#1}
\fi
\begin{document}
\bstctlcite{R1BibControl}

\title{FSC-Net: Integrating Fast Fourier Convolutions and Progressive Learning for Speech Bandwidth Extension}
\author{Xinan Chen, Xiaobin Rong, Qinwen Hu, Kai Chen, \IEEEmembership{Member, IEEE}, and Jing Lu, \IEEEmembership{Senior Member, IEEE}
\thanks{Manuscript received xxxxxxx xx, 202X; revised xxxxxxx xx, 202X; accepted xxxxxxx xx, 202X. Date of publication xxx xx, 202X; date of current version xxx xx, 202X. This work was supported by the National Natural Science Foundation of China under Grant 12274221. The associate editor coordinating the review of this manuscript and approving it for publication was xxx. (Corresponding author: Kai Chen.)}
\thanks{Xinan Chen, Xiaobin Rong, Qinwen Hu and Jing Lu are with the Key Laboratory of Modern Acoustics, Nanjing University, Nanjing 210008, China, and also with NJU-Horizon Intelligent Audio Lab, Horizon Robotics, Beijing 100094, China (e-mails: \{xinan.chen, xiaobin.rong, qinwen.hu\}@smail.nju.edu.cn; lujing@nju.edu.cn).}
\thanks{Kai Chen is with the Key Laboratory of Modern Acoustics, Nanjing University, Nanjing 210008, China (e-mail: chenkai@nju.edu.cn).}
\thanks{The source code will be released upon acceptance of the paper. Demo: \protect\url{https://xinan-chen.github.io/FSC-Net-demo}.}}

\markboth{IEEE Signal Processing Letters,~Vol.~XX, No.~XX,~202X}%
{Chen \MakeLowercase{\textit{et al.}}: FSC-Net: Integrating Fast Fourier Convolutions and Progressive Learning}
\maketitle

\begin{abstract}
Speech bandwidth extension (BWE) aims to reconstruct high-fidelity wideband audio from narrowband inputs. In this paper, we propose \textbf{FSC-Net} (Full-Spectrum Context Network), a parameter-efficient architecture designed to explicitly model cross-band harmonic dependencies. By integrating Fast Fourier Convolutions (FFCs) into a complex spectral mapping framework, FSC-Net expands its receptive field to the entire spectrum, capturing long-range frequency interactions effectively. To address the ill-posed nature of high-frequency generation, our novel frequency-progressive learning curriculum guides the network to reconstruct spectral details from coarse to fine. Experimental results on the VCTK and unseen EARS datasets demonstrate that FSC-Net delivers consistently strong reconstruction quality and generalization, particularly in the challenging VCTK 4~kHz-to-48~kHz task. Compared \rev{with} scaled-up baselines, our model \rev{achieves strong LSD, HF-LSD, NISQA, and ViSQOL results with only 1.54~M parameters.}
\end{abstract}

\begin{IEEEkeywords}
speech bandwidth extension, generative adversarial network, progressive learning, audio super-resolution
\end{IEEEkeywords}

\IEEEpeerreviewmaketitle

\section{Introduction}

Speech bandwidth extension (BWE) and audio super-resolution are fundamentally equivalent problems, both aiming to recover missing high-frequency spectral components. In practical speech communication systems, bandwidth limitations often degrade audio quality by truncating high-frequency components. BWE addresses this by reconstructing missing high-frequency content from narrowband signals \cite{iser2008bandwidth}. Early approaches relied on signal processing techniques including source-filter models \cite{makhoul1979high}, Line Spectral Frequencies (LSF) \cite{chennoukh2001speech}, codebook mapping \cite{carl1994bandwidth,unno2005robust}, and statistical methods using \rev{Gaussian mixture models (GMMs) and hidden Markov models (HMMs)} \cite{jax2003artificial,park2000narrowband}, though these often produced over-smoothed spectral parameters \cite{ling2015deep}.

Recent deep learning methods fall into two categories: waveform-based approaches that directly map narrowband to wideband waveforms \cite{kuleshov2017audio,li2021real,kim2019bandwidth,su2021bandwidth,han2022nu}, and spectrum-based methods that predict high-frequency amplitudes while estimating \rev{missing band phase \rOne{from low-band phase patterns}} \cite{li2015deep,abel2018simple} or vocoders \cite{botinhao2006frequency}. Alternative approaches attempt phase recovery through the \rev{short-time Fourier transform (STFT)} \cite{baenet,AP-BWE} or \rev{modified discrete cosine transform (MDCT)} spectra \cite{shuai2023mdctgan}, but phase reconstruction remains challenging. \rev{Diffusion-based generative approaches probabilistically restore ambiguous content but require iterative sampling \cite{lemercier2023analysing,lemercier2025review}; FSC-Net instead performs single-pass spectral mapping.}


For instance, AERO \cite{aero} incurs high computational costs and often produces metallic artifacts in the reconstructed high frequencies.
Similarly, AP-BWE \cite{AP-BWE} suffers from parameter explosion ($\sim$30~M) and \rOne{a large memory footprint} due to stacked \rev{fully connected (FC)} layers. Conversely, BAE-Net \cite{baenet} achieves efficiency via compressed FCs but lacks the capacity and global context to capture complex spectral dependencies. Even when scaled up with more parameters, such architectures often hit a performance bottleneck due to their limited receptive fields and lack of global context modeling. Recently, SFNet \cite{dai2025sfnet} introduced a highly efficient neural source-filter framework by integrating traditional digital signal processing (DSP) modules. \rev{However, it depends on explicit pitch tracking, whereas FSC-Net performs direct spectral mapping.}

These limitations motivate our improved approach. Inspired by advances in image restoration \cite{lama}, we propose FSC-Net (Full-Spectrum Context Network), which integrates Fast Fourier Convolutions (FFCs) into the TF-GridNet backbone. \rev{FFC-SE previously adapted FFC to complex-spectrogram speech enhancement via autoencoder and U-Net architectures \cite{shchekotov2022ffcse}; FSC-Net instead embeds FFC in TF-GridNet for BWE and combines it with residual-based frequency-progressive supervision.}

\begin{figure*}[t]
\centerline{\includegraphics[width=\textwidth]{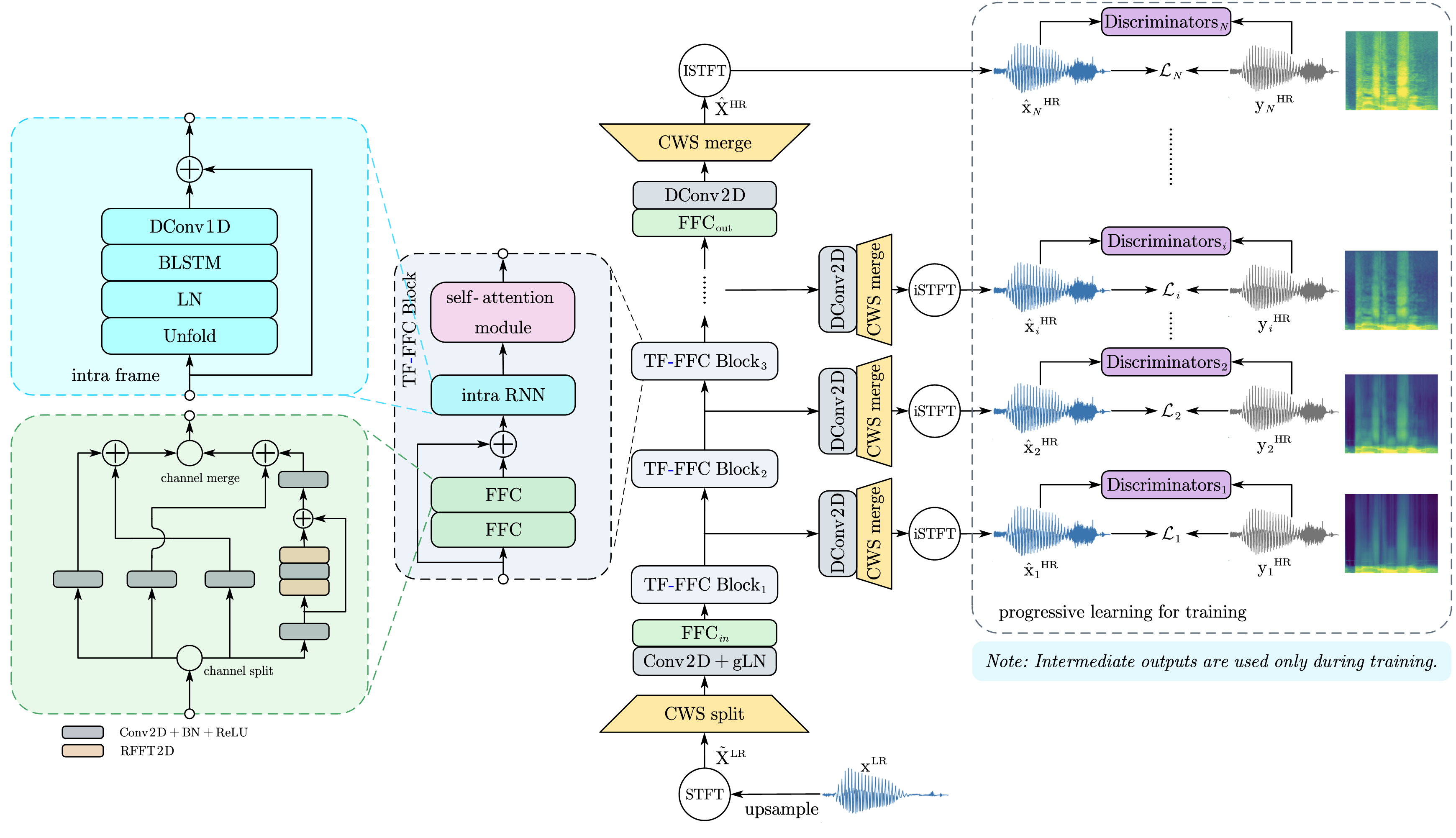}}
\vspace{-0.1cm}
\caption{The general architecture of FSC-Net and the framework for progressive learning. \rev{The left panels detail the intra-frame and FFC modules, while the right panel shows the stage-wise outputs and discriminators used only during training.}
}
\label{main}
\vspace{-0.4cm}
\end{figure*}

Beyond architectural augmentation, we introduce a frequency-progressive learning paradigm. Unlike existing \rev{signal-to-noise ratio (SNR)}/\rev{signal-to-interference ratio (SIR)}-progressive strategies for noise suppression \cite{hou2024snr, gao2016snr}, \rev{we construct progressively refined targets by frequency-axis smoothing of high-resolution residual spectrograms. This coarse-to-fine pathway guides the model from the global spectral envelope to harmonic details, mitigating artificial harmonics and spectral oversmoothing }\cite{aero, jang2021univnet}. \rev{FSC-Net achieves favorable objective results for high-frequency reconstruction, particularly in the challenging 4~kHz~$\to$~48~kHz setting.}

\section{Methodology}

\subsection{Problem Formulation}
The \rev{aim} of BWE is to estimate a high-resolution wideband waveform $\mathbf{y}^{\mathrm{HR}} \in \mathbb{R}^{T_{\mathrm{wb}}}$ from a given narrowband input waveform $\mathbf{x}^{\mathrm{LR}} \in \mathbb{R}^{T_{\mathrm{nb}}}$, \rev{where $T_{\mathrm{wb}}$ and $T_{\mathrm{nb}}$ denote the waveform lengths at their respective sampling rates}. \rev{We first upsample $\mathbf{x}^{\mathrm{LR}}$ \rOne{to the target sampling rate, obtaining} $\tilde{\mathbf{x}}^{\mathrm{LR}}$, and compute $\tilde{X}^{\mathrm{LR}}=\operatorname{STFT}(\tilde{\mathbf{x}}^{\mathrm{LR}}) \in \mathbb{C}^{T \times F_{\mathrm{wb}}}$.}
The proposed FSC-Net acts as a complex spectral mapper $\mathcal{G}_\theta$, which takes \rev{$\tilde{X}^{\mathrm{LR}}$} as input and predicts the full-band complex spectrogram $\hat{X}^{\mathrm{HR}} \in \mathbb{C}^{T \times F_{\mathrm{wb}}}$:
\begin{equation}
    \rev{\hat{X}^{\mathrm{HR}} = \mathcal{G}_\theta(\tilde{X}^{\mathrm{LR}}).}
\end{equation}
The predicted wideband spectrogram is then transformed back to the time domain through the \rev{inverse STFT (iSTFT)} to obtain the predicted waveform $\hat{\mathbf{x}}^{\mathrm{HR}} = \text{iSTFT}(\hat{X}^{\mathrm{HR}})$, as illustrated in Fig.~\ref{main}.

\subsection{FSC-Net Architecture}
Our proposed model is built upon the TF-GridNet \cite{tfgridnet}, a state-of-the-art time-frequency domain model originally designed for speech separation. We adapt and enhance this architecture for the BWE task by introducing the following two key components.

\subsubsection{Channel-wise Subband (CWS) Processing}
 Given an input complex spectrogram $X \in \mathbb{C}^{T \times F}$, we partition the frequency dimension $F$ into $B$ subbands. These subbands are then stacked along the channel dimension, transforming the input into a tensor $X_{\mathrm{cws}} \in \mathbb{C}^{B \times T \times \rev{ \left\lfloor F/B \right\rfloor}}$. This CWS operation \cite{rong2025ts} allows the model to capture inter-subband dependencies through channel interactions while reducing the computational burden on the frequency axis.

\subsubsection{Fast Fourier Convolution (FFC) Integration}
To expand the receptive field for recovering missing high-frequency components correlated with low-frequency content, we replace TF-GridNet's time-domain inter-RNN with FFC modules \cite{lama}, retaining the intra-RNN. 
The FFC splits the input feature map into two branches: a local branch using standard convolutions and a global branch using spectral transforms. The global branch applies \rOne{a real-valued fast Fourier transform (RFFT)} to the feature map, performs convolutions in the frequency domain, and then applies \rOne{an inverse real FFT (IRFFT)}. Formally, for an input feature map $X$ the FFC output $Y$ is defined as:
\begin{equation}
    Y = \text{Conv2D}_{\mathrm{local}}(X) + \rOne{\text{IRFFT}}(\text{Conv2D}_{\mathrm{global}}(\rOne{\text{RFFT}}(X))).
\end{equation}
This mechanism provides the model with a global receptive field covering the entire spectrum, effectively bridging the ``spectral gap'' in BWE tasks.

\subsection{Progressive Learning Strategy}

\rev{Kim et al. progressively supervised intermediate waveforms at increasing sampling rates using multi-resolution reconstruction losses and multiple discriminators~\cite{kim2021multiresolution}; FSC-Net similarly employs these training components, but all stages operate at a fixed target rate using frequency-smoothed magnitude-residual targets.} Directly hallucinating high-frequency details in a single step often causes metallic artifacts \cite{aero}. To address this, we propose a coarse-to-fine frequency-progressive learning strategy (Fig.~\ref{main}), supervising each of the $N$ TF-FFC Blocks with a progressive target $|Y_{i}|$. Rather than smoothing the target directly, we apply sliding-window averaging to the magnitude residual $R(t, f) = |Y^{\mathrm{HR}}(t, f)| - |\rev{\tilde{X}^{\mathrm{LR}}}(t, f)|$ between the ground-truth and input spectrograms. For the $i$-th stage, the target is computed as:
\begin{equation}
    |Y_{i}(t, f)| = |\rev{\tilde{X}^{\mathrm{LR}}}(t, f)| + \frac{1}{W_i} \sum_{\delta=-\left\lfloor W_i/2 \right\rfloor}^{\left\lfloor W_i/2 \right\rfloor} R(t, f+\delta),
\end{equation}
where $W_i \in \{257, 65, 17, 5, 1\}$ is a decreasing sequence of window sizes, with zero-padding at frequency boundaries. Early blocks use larger $W_i$ to learn the global spectral envelope of missing components, while deeper blocks use smaller $W_i$ to refine fine harmonic structures. Ultimately, $W_N = 1$ strictly recovers the exact ground truth ($|Y_{N}| = |Y^{\mathrm{HR}}|$).

\subsection{Loss Functions}
We define the Multi-Resolution STFT loss as the average of a spectral convergence term $\mathcal{L}_{\mathrm{sc}}$ and a log-magnitude term $\mathcal{L}_{\mathrm{mag}}$ over $M$ FFT resolutions:
\begin{align}
    \mathcal{L}_{\mathrm{sc}}(\hat{X}, Y) &= \frac{1}{M}\sum_{m=1}^{M} \frac{\bigl\| |Y_m| - |\hat{X}_m| \bigr\|_{\mathrm{F}}}{\bigl\| |Y_m| \bigr\|_{\mathrm{F}}}, \\
    \mathcal{L}_{\mathrm{mag}}(\hat{X}, Y) &= \frac{1}{M}\sum_{m=1}^{M} \left\| \log \frac{|Y_m|}{|\hat{X}_m|} \right\|_1, \\
    \mathcal{L}_{\mathrm{mr\text{-}stft}}(\hat{X}, Y) &= \mathcal{L}_{\mathrm{sc}}(\hat{X}, Y) + \mathcal{L}_{\mathrm{mag}}(\hat{X}, Y),
\end{align}
where $\|\cdot\|_{\mathrm{F}}$ is the Frobenius norm,  $\|\cdot\|_{\mathrm{1}}$ is the 1‑norm, and the subscript $m$ indexes the $m$-th FFT resolution. The Log-Spectral Distance is
\begin{equation}
    \mathcal{L}_{\mathrm{lsd}}(\hat{X}, Y) = \mathbb{E} \left[ \sqrt{ \frac{1}{F} \sum_{f} \left( \log_{10} \frac{|Y|^2 + \epsilon}{|\hat{X}|^2 + \epsilon} \right)^2 } \right],
\end{equation}
with $\epsilon$ preventing numerical instability. For the $i$-th stage we substitute $\hat{X} \leftarrow \hat{X}_i$, $Y \leftarrow Y_{i}$ and aggregate:
\begin{equation}
    \mathcal{L}_i = \mathcal{L}_{\mathrm{mr\text{-}stft}}(\hat{X}_i, Y_{i}) + \lambda_{\mathrm{lsd}}\,\mathcal{L}_{\mathrm{lsd}}(\hat{X}_i, Y_{i}),
\end{equation}
which corresponds to $\mathcal{L}_1, \dots, \mathcal{L}_N$ in Fig.~\ref{main}.

For stable adversarial training, we adopt the Least Squares GAN \cite{mao2017least} framework, applied to every stage output \rev{$\hat{\mathbf{x}}_i = \text{iSTFT}(\hat{X}_i)$}. Each stage $i$ is paired with its own multi-scale discriminator $\mathcal{D}_{i}$ \cite{melgan} comprising $K$ sub-discriminators $\{\mathcal{D}_{i,j}\}_{j=1}^{K}$, \rev{whose $l$-th layer output is denoted $\mathcal{D}_{i,j}^{(l)}$. The adversarial and feature-matching losses are:}
\begin{align}
    \mathcal{L}_{\mathrm{adv}} &= \frac{1}{N K} \sum_{i,j} \mathbb{E} \bigl[ (1 - \mathcal{D}_{i,j}(\rev{\hat{\mathbf{x}}_i}))^2 \bigr], \\
    \mathcal{L}_{\mathrm{feat}} &= \frac{1}{N K} \sum_{i,j} \mathbb{E}_{l} \bigl[ \bigl\| \mathcal{D}_{i,j}^{(l)}(\rev{\mathbf{y}^{\mathrm{HR}}}) - \mathcal{D}_{i,j}^{(l)}(\rev{\hat{\mathbf{x}}_i}) \bigr\|_1 \bigr].
\end{align}

The generator $\mathcal{G}$ and the per-stage discriminators $\{\mathcal{D}_{i}\}_{i=1}^{N}$ are trained jointly using the overall objective:
\vspace{-0.1cm}
\begin{equation}
    \mathcal{L}_{G} = \sum_{i=1}^{N} \mathcal{L}_i + \lambda_{\mathrm{adv}} (\mathcal{L}_{\mathrm{adv}} + \lambda_{\mathrm{feat}} \mathcal{L}_{\mathrm{feat}}).
\end{equation}
\vspace{-0.2cm}
\section{Experiments}

\subsection{Experimental Setup}
We evaluate our method on the VCTK corpus (version 0.92)~\cite{vctk}, partitioned into 100 training speakers and the last 8 speakers held out for testing (no speaker overlap). The 48~kHz recordings serve as the high-resolution targets, while narrowband inputs are obtained by downsampling to 4~kHz or 16~kHz \rev{using torchaudio resampling with its default anti-aliasing low-pass filter}. For evaluation, we adopt Log-Spectral Distance (LSD), \rev{high-frequency LSD (HF-LSD), Non-Intrusive Speech Quality Assessment (NISQA)}~\cite{NISQA}, \rev{Virtual Speech Quality Objective Listener (ViSQOL) \cite{visqol}, and Perceptual Evaluation of Speech Quality (PESQ)~\cite{PESQ}~\cite{itut_p8622_2007}.} \rev{HF-LSD is restricted to the missing band above the input Nyquist frequency (2~kHz or 8~kHz), while ViSQOL is evaluated in audio mode at 48~kHz}. \rev{PESQ is computed after aligning and resampling both signals to 16~kHz, and is reported only for 4~kHz~$\to$~48~kHz, since its 8~kHz upper limit excludes the entire reconstructed band in the 16~kHz case. Thus, for 4~kHz input, PESQ evaluates only the added 2--8~kHz band; as a full-reference metric, it tends to reward conservative rather than generative reconstruction.}

\subsection{Implementation Details}
During training, audio segments are randomly cropped to a fixed length of 2~s. Our enhanced TF-GridNet is configured with $N=5$ blocks and $B=3$ subbands. We employ a 32~ms window and 16~ms hop size for STFT at 48~kHz. The learning rate follows a warm-up and cosine decay schedule, peaking at $5 \times 10^{-4}$ for the generator and $2.6 \times 10^{-4}$ for the discriminator. \rOne{In our experiments, we set} $\lambda_{\mathrm{lsd}} = 5$, 
$\lambda_{\mathrm{adv}} = 0.34$, and $\lambda_{\mathrm{feat}} = 0.1$. 
\rev{Model and training hyperparameters were selected on the validation set.} \rOne{For the models profiled in this work (Tables~\ref{tab:comparison_unified} and~\ref{tab:ablation}), parameter and MAC counts cover only modules used at inference. Both counts exclude training-only components, including discriminators and auxiliary output branches used for intermediate supervision. MACs were estimated with THOP for 1-s inputs.}

\subsection{Comparison with State-of-the-Art Methods}
\rOne{We compare FSC-Net with} several competitive baselines: AP-BWE \cite{AP-BWE}, BAE-Net \cite{baenet}, AERO \cite{aero}, and the recently proposed SFNet \cite{dai2025sfnet}\footnote{The results of SFNet are directly cited from the original paper \cite{dai2025sfnet}, where NISQA and PESQ were not reported. \rev{All other models are trained with the same configuration and losses.}}. To ensure a fair comparison, particularly against the lightweight BAE-Net, we introduced a scaled-up version denoted as BAE-Net*\footnote{We scaled up the BAE-Net by increasing the channel dimensions and network depth to match the computational scale (MACs) of typical high-performance models, \rev{thereby enabling comparison at a similar computational budget.}}.

\begin{table*}[t]
\vspace{-0.3cm}
\caption{Performance comparison on VCTK for 4~kHz~$\to$~48~kHz and 16~kHz~$\to$~48~kHz BWE.}
\label{tab:comparison_unified}
\centering
\footnotesize
\setlength{\tabcolsep}{4pt}
\renewcommand{\arraystretch}{1.1}
\begin{tabular}{l|ccccc|cccc|cc}
\hline
\multirow{2}{*}{\rev{\textbf{Method}}} & \multicolumn{5}{c|}{\textbf{4~kHz~$\to$~48~kHz}} & \multicolumn{4}{c|}{\textbf{16~kHz~$\to$~48~kHz}} & \multirow{2}{*}{\textbf{Params (M)}} & \multirow{2}{*}{\textbf{MACs (G)}} \\
\cline{2-6}\cline{7-10}
& \textbf{LSD}$\downarrow$ & \rev{\textbf{HF-LSD}$\downarrow$} & \textbf{PESQ}$\uparrow$ & \textbf{NISQA}$\uparrow$ & \rev{\textbf{ViSQOL}$\uparrow$} & \textbf{LSD}$\downarrow$ & \rev{\textbf{HF-LSD}$\downarrow$} & \textbf{NISQA}$\uparrow$ & \rev{\textbf{ViSQOL}$\uparrow$} & & \\
\hline
\rev{Input} & \rev{3.688} & \rev{3.852} & \rev{3.057} & \rev{2.978} & \rev{2.065} & \rev{2.954} & \rev{3.612} & \rev{4.293} & \rev{2.260} & \rev{--} & \rev{--} \\
\hline
AP-BWE \cite{AP-BWE} & 0.955 & \rev{1.023} & \rev{2.792} & 4.256 & \rev{3.047} & 0.729 & \rev{0.864} & 4.391 & \rev{3.499} & 29.76 & 17.87 \\
BAE-Net lite \cite{baenet} & 0.989 & \rev{1.028} & \rev{2.809} & 4.142 & \rev{2.977} & 0.722 & \rev{0.891} & 4.312 & \rev{3.383} & \textbf{0.57} & \textbf{0.057} \\
BAE-Net* & 0.904 & \rev{0.937} & \rev{3.008} & 4.221 & \rev{3.109} & 0.714 & \rev{0.867} & \textbf{4.503} & \rev{3.505} & 17.41 & 26.32 \\
AERO \cite{aero} & 0.992 & \rev{1.017} & \rev{2.527} & 4.280 & \rev{3.092} & 0.789 & \rev{0.879} & 4.267 & \rev{3.436} & 21.66 & 51.74 \\
SFNet \cite{dai2025sfnet} & 0.920 & -- & -- & -- & \rev{3.230} & 0.730 & -- & -- & \rev{\textbf{3.610}} & 1.33 & 0.88 \\
\textbf{FSC-Net} & \textbf{0.877} & \rev{\textbf{0.908}} & \rev{\textbf{3.082}} & \textbf{4.313} & \rev{\textbf{3.256}} & \textbf{0.705} & \rev{\textbf{0.836}} & 4.468 & \rev{3.549} & 1.54 & 27.74 \\
\hline
\end{tabular}
\vspace{-0.6cm}
\end{table*}

\subsubsection{Performance on VCTK}
\rev{As shown in Table \ref{tab:comparison_unified}, FSC-Net performs best across all metrics for 4~kHz~$\to$~48~kHz. For 16~kHz~$\to$~48~kHz, FSC-Net achieves the lowest LSD and HF-LSD while remaining competitive on NISQA and ViSQOL. Notably, under identical training objectives and protocols, FSC-Net outperforms BAE-Net* on eight of the nine reported metrics while using substantially fewer parameters (1.54~M vs.\ 17.41~M), demonstrating its stronger overall performance and parameter efficiency. \rOne{FSC-Net requires 27.74~G MACs for a 1-s input, exceeding the cost of several baselines in Table~\ref{tab:comparison_unified}.}}

\subsubsection{Generalization to Unseen EARS Dataset}
To evaluate the generalization capability of the proposed model, we conducted zero-shot testing on \rev{the full EARS dataset} \cite{ears} (16~kHz~$\to$~48~kHz) without any fine-tuning.

\begin{table}[htbp]
\vspace{-0.2cm}
\caption{Generalization Performance on EARS Dataset (16~kHz~$\to$~48~kHz)}
\label{tab:ears}
\centering
\footnotesize
\setlength{\tabcolsep}{6pt}
\renewcommand{\arraystretch}{1.1}
\begin{tabular}{l|cccc}
\hline
\rev{\textbf{Method}} & \textbf{LSD}$\downarrow$ & \rev{\textbf{HF-LSD}$\downarrow$} & \textbf{NISQA}$\uparrow$ & \rev{\textbf{ViSQOL}$\uparrow$} \\
\hline
\rev{Input} & \rev{1.626} & \rev{1.980} & \rev{3.709} & \rev{3.394} \\
\hline
AP-BWE \cite{AP-BWE} & 1.425 & \rev{1.703} & 3.615 & \rev{3.271} \\
BAE-Net lite \cite{baenet} & 1.326 & \rev{1.560} & 3.817 & \rev{3.597} \\
BAE-Net* & 1.224 & \rev{1.468} & 3.802 & \rev{\textbf{3.683}} \\
AERO \cite{aero} & 1.280 & \rev{1.507} & 3.825 & \rev{3.410} \\
\textbf{FSC-Net} & \textbf{1.207} & \rev{\textbf{1.398}} & \textbf{3.921} & \rev{3.663} \\
\hline
\end{tabular}
\vspace{-0.2cm}
\end{table}

\rev{As shown in Table \ref{tab:ears}, FSC-Net achieves the best LSD, HF-LSD, and NISQA while remaining competitive on ViSQOL, supporting its robustness to unseen speakers and recording conditions.}

\subsection{Qualitative Analysis}
\begin{figure}[t!] 
    \vspace{-0.15cm}
    \centering
    
    \subfloat[Target $\text{LSD}=0.0000$]{\includegraphics[width=0.48\columnwidth, trim=5 5 5 5, clip]{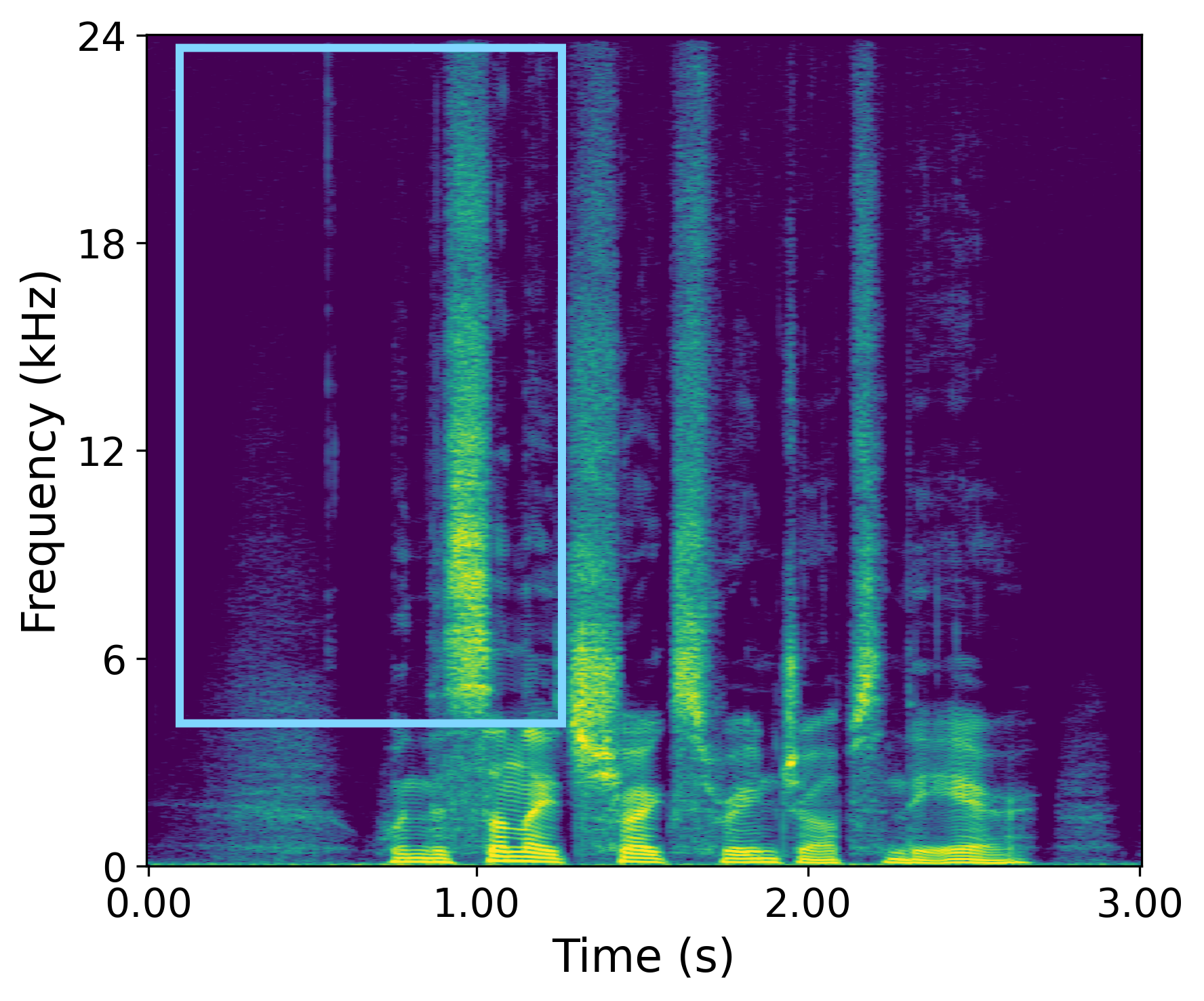}}
    \hfill
    \subfloat[Input $\text{LSD}=3.5918$]{\includegraphics[width=0.48\columnwidth, trim=5 5 5 5, clip]{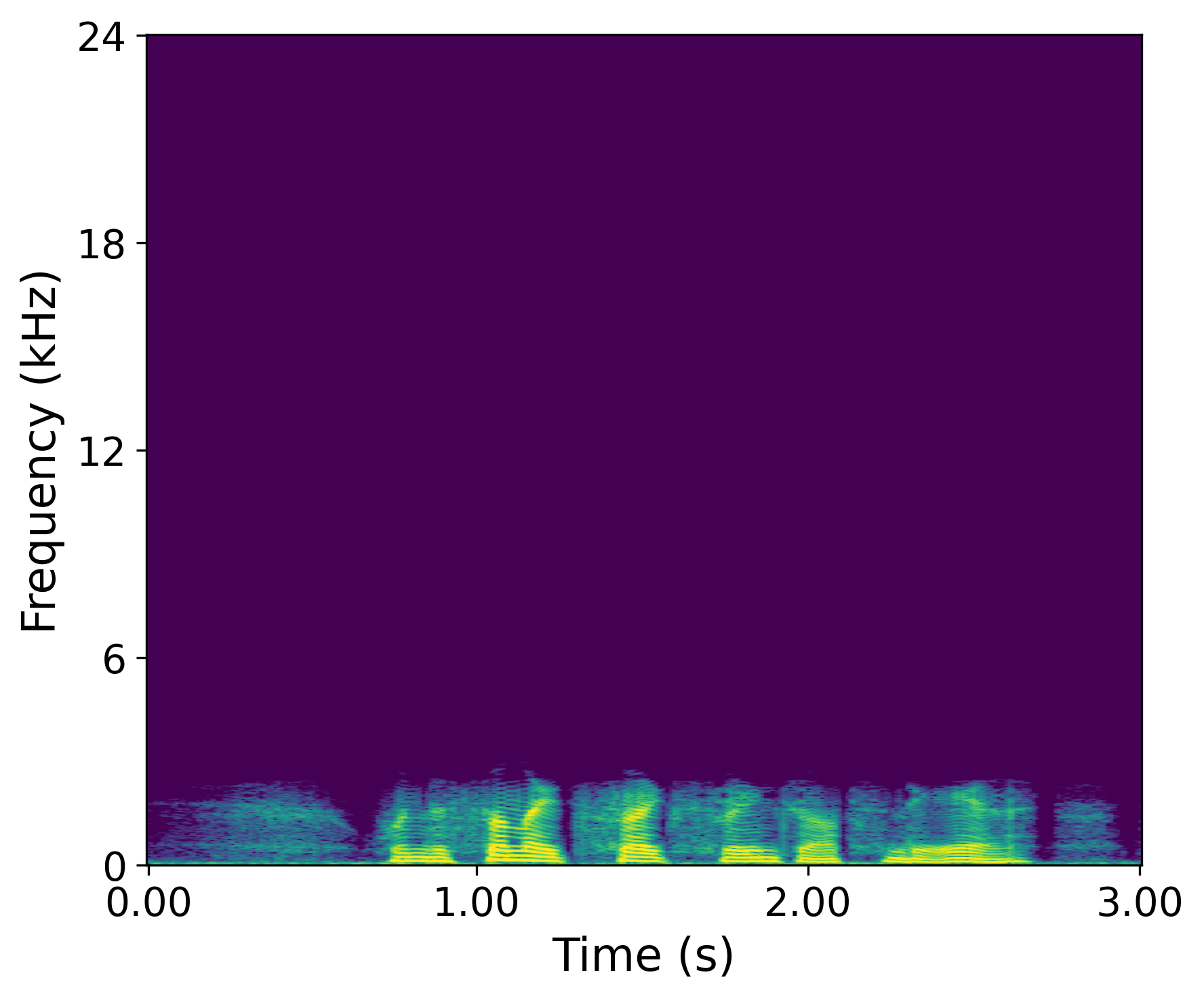}}
    
    \vspace{-0.3cm} 
    
    \subfloat[FSC-Net $\text{LSD}=0.8863$]{\includegraphics[width=0.48\columnwidth, trim=5 5 5 5, clip]{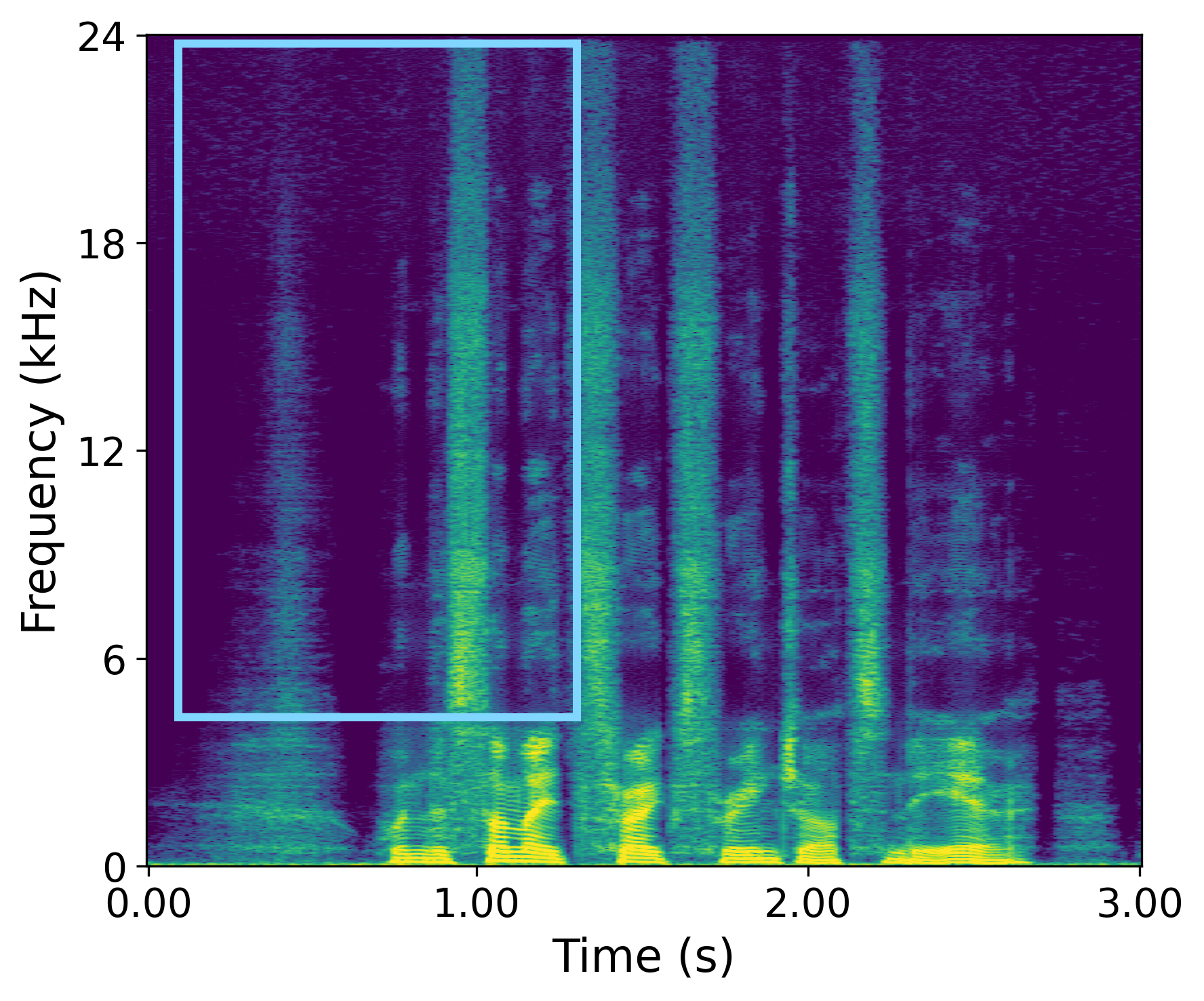}}
    \hfill
    \subfloat[AP-BWE $\text{LSD}=1.0929$]{\includegraphics[width=0.48\columnwidth, trim=5 5 5 5, clip]{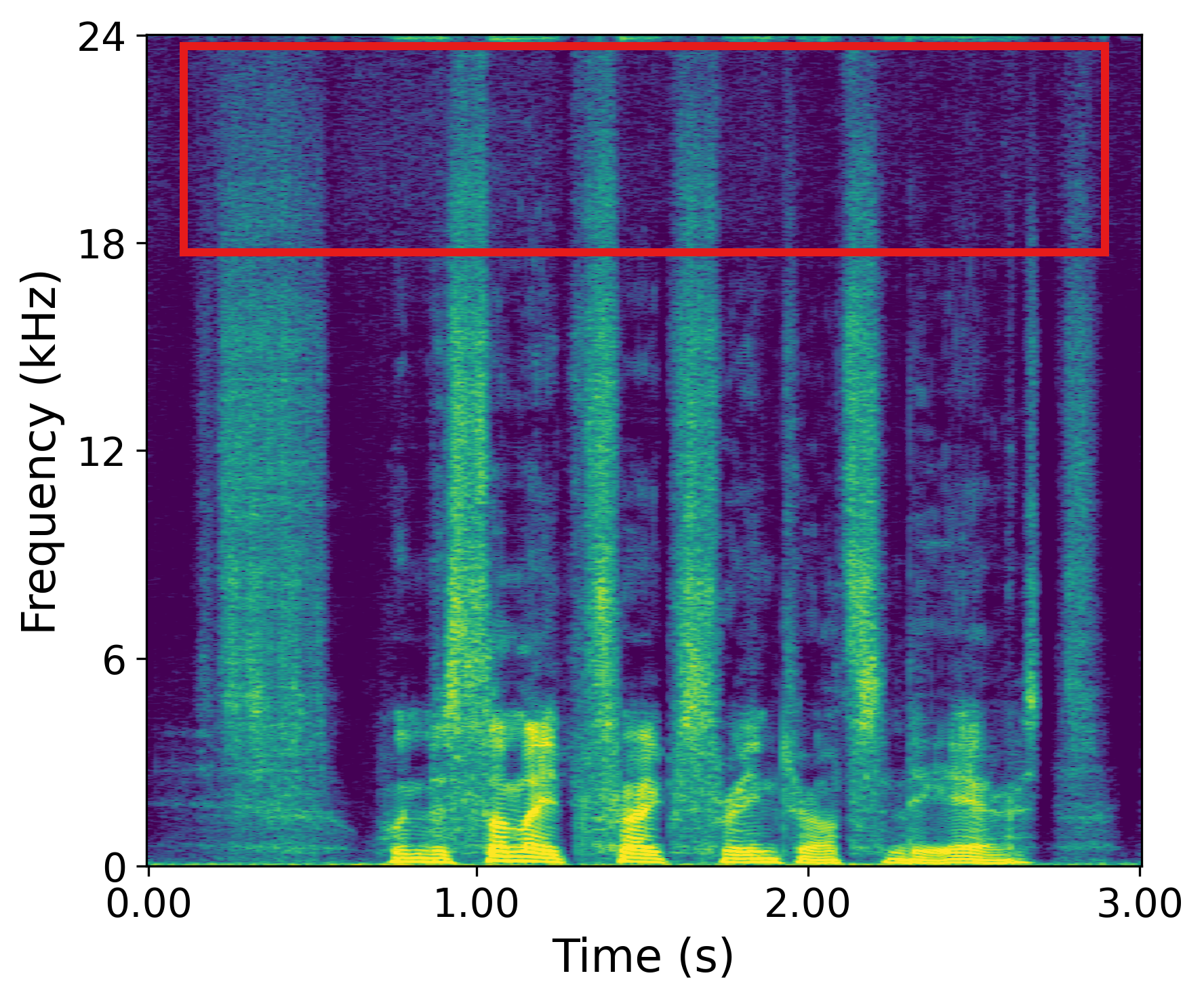}}
    
    \vspace{-0.3cm}
    
    \subfloat[BAE-Net $\text{LSD}=0.9953$]{\includegraphics[width=0.48\columnwidth, trim=5 5 5 5, clip]{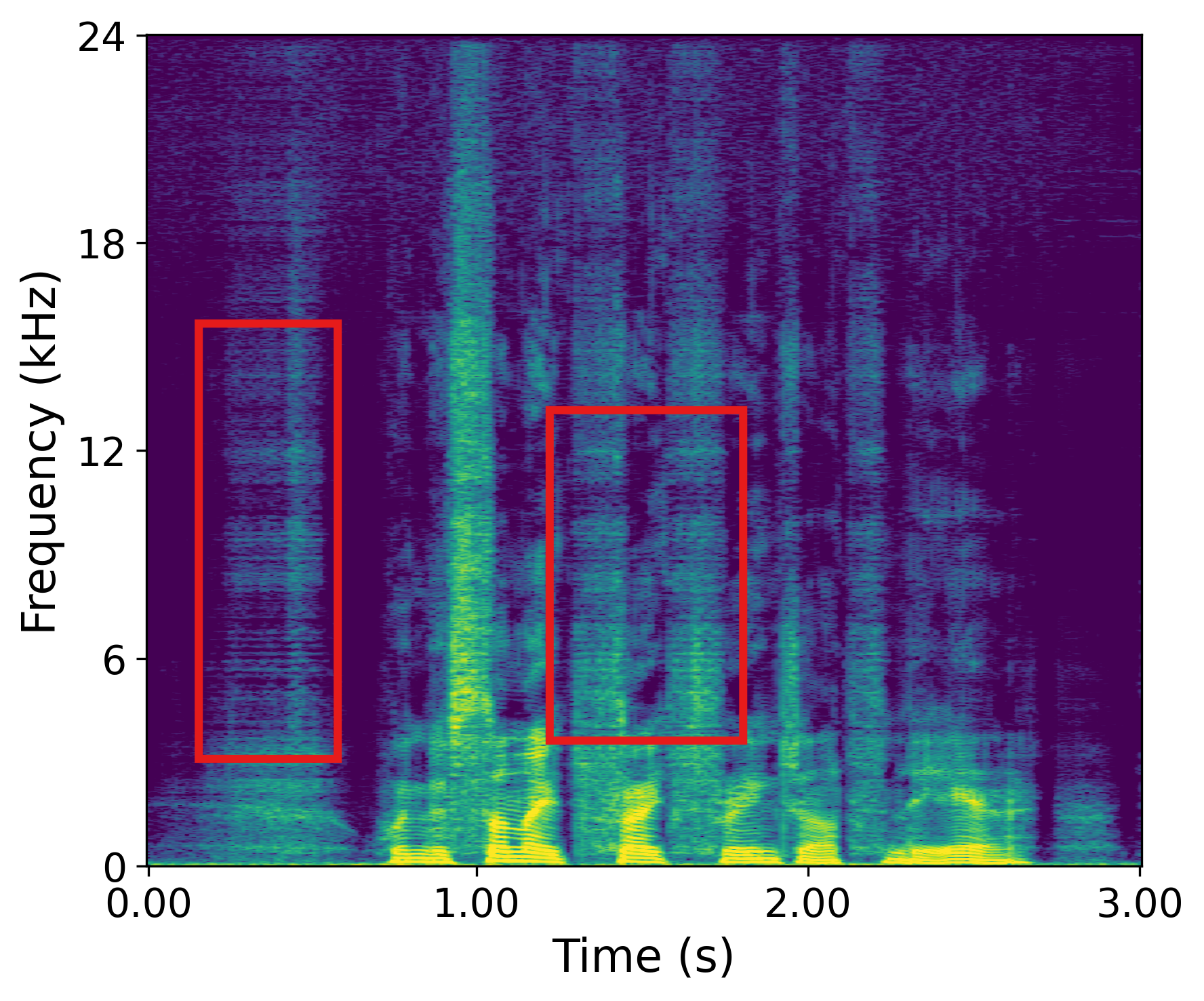}}
    \hfill
    \subfloat[AERO $\text{LSD}=0.9168$]{\includegraphics[width=0.48\columnwidth, trim=5 5 5 5, clip]{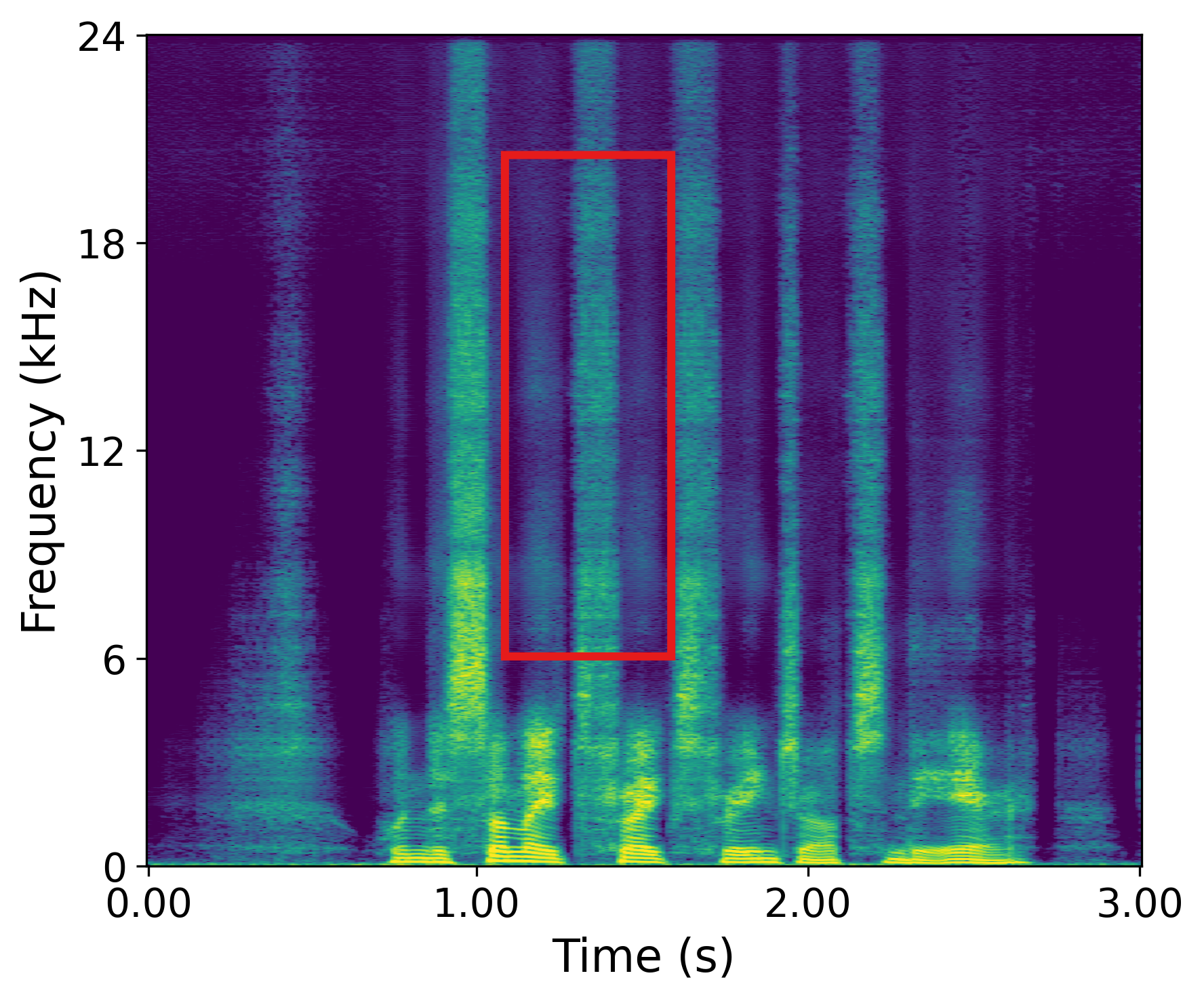}}
    
    \vspace{-0.1cm}
    \caption{\rev{Spectrograms of the ground-truth speech and the speech signals extended by baseline methods and FSC-Net on VCTK, with input and target sampling rates of 4~kHz and 48~kHz, respectively. Red boxes mark baseline artifacts; blue boxes mark the corresponding regions reconstructed by FSC-Net.}}

    \label{fig:spec}
    
    \vspace{-0.4cm}
\end{figure}

Fig. \ref{fig:spec} visualizes the reconstructed spectrograms. As highlighted by the \textbf{red boxes}, baseline methods suffer from distinct artifacts: excessive noise (AP-BWE), energy discontinuities (BAE-Net), and \rev{artificial horizontal striations with over-smoothed textures} (AERO). Conversely, \textbf{FSC-Net} (\textbf{blue boxes}) accurately restores continuous and clear harmonic structures akin to the ground truth. Benefiting from FFC's global context and the progressive curriculum, our model successfully suppresses these spectral anomalies. \rev{In the illustrated example, FSC-Net exhibits more continuous high-frequency harmonic patterns with fewer visible spectral discontinuities than the evaluated baselines.}

\subsection{Ablation Study}
We conducted an ablation study to verify the contribution of each proposed component. \rev{All configurations use identical loss functions to isolate the effects of FFC and progressive learning (PL).} The results are summarized in Table \ref{tab:ablation}.

\begin{table}[htbp]
\caption{Ablation Study on VCTK (4~kHz~$\to$~48~kHz)}
\label{tab:ablation}
\centering
\scriptsize
\setlength{\tabcolsep}{1pt}
\renewcommand{\arraystretch}{1.1}
\begin{tabular}{l|cccccc}
\hline
\rev{\textbf{Method}} & \textbf{LSD}$\downarrow$ & \rev{\textbf{HF-LSD}$\downarrow$} & \textbf{PESQ}$\uparrow$ & \textbf{NISQA}$\uparrow$ & \rev{\textbf{ViSQOL}$\uparrow$} & \rev{\textbf{Par. (M) / MACs (G)}} \\
\hline
\rev{w/o FFC \& PL} & \rev{0.919} & \rev{0.956} & \rev{2.763} & \rev{4.140} & \rev{3.082}& \rev{2.77 / 49.68}  \\
\rev{w/o PL} & \rev{0.886} & \rev{0.919} & \rev{3.054} & \rev{4.205} & \rev{3.153} & \rev{1.54 / 27.74}\\
\textbf{FSC-Net} & \textbf{0.877} & \rev{\textbf{0.908}} & \rev{\textbf{3.082}} & \textbf{4.313} & \rev{\textbf{3.256}} & \rev{1.54 / 27.74}\\
\hline
\end{tabular}
\vspace{-0.1cm}
\end{table}

\rev{FFC improves all metrics over the model without FFC and PL. Adding PL further reduces HF-LSD from 0.919 to 0.908 and improves NISQA and ViSQOL from 4.205 and 3.153 to 4.313 and 3.256, respectively.}

\vspace{-0.15cm}
\section{Conclusion}
\rOne{We propose FSC-Net for high-fidelity speech bandwidth extension, combining Fast Fourier Convolutions to model cross-band dependencies with frequency-progressive learning to recover high-frequency detail. On VCTK and in zero-shot tests on EARS, FSC-Net (1.54~M parameters) delivers consistent HF-LSD gains and competitive ViSQOL and NISQA scores against leading baselines, including scaled-up models.}

\bibliographystyle{IEEEtran}
\bibliography{R1_ref,R1_bibcontrol}

@IEEEtranBSTCTL{R1BibControl,
  CTLdash_repeated_names = "no"
}

@article{AP-BWE,
  title={Towards high-quality and efficient speech bandwidth extension with parallel amplitude and phase prediction},
  author={Lu, Ye-Xin and Ai, Yang and Du, Hui-Peng and Ling, Zhenhua},
  journal={IEEE/ACM Trans. Audio, Speech, Lang. Process.},
  volume={33},
  pages={236--250},
  year={\rOne{2025}}
}

@inproceedings{baenet,
  title={{BAE-Net}: A low complexity and high fidelity bandwidth-adaptive neural network for speech super-resolution},
  author={Yu, Guochen and Zheng, Xiguang and Li, Nan and Han, Runqiang and Zheng, Chengshi and Zhang, Chen and Zhou, Chao and Huang, Qi and Yu, Bing},
  booktitle={Proc. IEEE Int. Conf. Acoust., Speech, Signal Process. (ICASSP)},
  pages={571--575},
  year={2024}
}

@inproceedings{aero,
  title={{AERO}: Audio super resolution in the spectral domain},
  author={Mandel, Moshe and Tal, Or and Adi, Yossi},
  booktitle={Proc. IEEE Int. Conf. Acoust., Speech, Signal Process. (ICASSP)},
  pages={1--5},
  year={2023}
}

@inproceedings{chennoukh2001speech,
  title={Speech enhancement via frequency bandwidth extension using line spectral frequencies},
  author={Chennoukh, Samir and Gerrits, A and Miet, G and Sluijter, R},
  booktitle={Proc. IEEE Int. Conf. Acoust., Speech, Signal Process. (ICASSP)},
  volume={1},
  pages={665--668},
  year={2001}
}

@inproceedings{makhoul1979high,
  title={High-frequency regeneration in speech coding systems},
  author={Makhoul, John and Berouti, Michael},
  booktitle={Proc. IEEE Int. Conf. Acoust., Speech, Signal Process. (ICASSP)},
  volume={4},
  pages={428--431},
  year={1979}
}

@book{iser2008bandwidth,
  title={Bandwidth extension of speech signals},
  author={Iser, Bernd and Minker, Wolfgang and Schmidt, Gerhard},
  year={2008},
  publisher={Springer}
}

@inproceedings{carl1994bandwidth,
  title={Bandwidth enhancement of narrowband speech signals},
  author={Carl, Holger},
  booktitle={Proc. Eur. Signal Process. Conf. (EUSIPCO)},
  volume={2},
  pages={1178--1181},
  year={1994}
}

@inproceedings{unno2005robust,
  title={A robust narrowband to wideband extension system featuring enhanced codebook mapping},
  author={Unno, Takahiro and McCree, Alan},
  booktitle={Proc. IEEE Int. Conf. Acoust., Speech, Signal Process. (ICASSP)},
  volume={1},
  pages={I-805--I-808},
  year={2005}
}

@inproceedings{jax2003artificial,
  title={Artificial bandwidth extension of speech signals using {MMSE} estimation based on a hidden {Markov} model},
  author={Jax, Peter and Vary, Peter},
  booktitle={Proc. IEEE Int. Conf. Acoust., Speech, Signal Process. (ICASSP)},
  volume={1},
  pages={I-680--I-683},
  year={2003},
  doi={10.1109/ICASSP.2003.1198872}
}

@inproceedings{park2000narrowband,
  title={Narrowband to wideband conversion of speech using {GMM} based transformation},
  author={Park, Kun-Youl and Kim, Hyung Soon},
  booktitle={Proc. IEEE Int. Conf. Acoust., Speech, Signal Process. (ICASSP)},
  volume={3},
  pages={1843--1846},
  year={2000}
}

@article{ling2015deep,
  title={Deep learning for acoustic modeling in parametric speech generation: A systematic review of existing techniques and future trends},
  author={Ling, Zhen-Hua and Kang, Shi-Yin and Zen, Heiga and Senior, Andrew and Schuster, Mike and Qian, Xiao-Jun and Meng, Helen M and Deng, Li},
  journal={IEEE Signal Process. Mag.},
  volume={32},
  number={3},
  pages={35--52},
  year={2015}
}

@article{tfgridnet,
  title={{TF-GridNet}: Integrating full- and sub-band modeling for speech separation},
  author={Wang, Zhong-Qiu and Cornell, Samuele and Choi, Shukjae and Lee, Younglo and Kim, Byeong-Yeol and Watanabe, Shinji},
  journal={IEEE/ACM Trans. Audio, Speech, Lang. Process.},
  volume={31},
  pages={3221--3236},
  year={2023}
}

@inproceedings{kuleshov2017audio,
  title={Audio super resolution using neural networks},
  author={Kuleshov, Volodymyr and Enam, S. Zayd and Ermon, Stefano},
  booktitle={Proc. Int. Conf. Learn. Represent. (ICLR) Workshop},
  year={2017}
}

@inproceedings{li2021real,
  title={Real-time speech frequency bandwidth extension},
  author={Li, Yunpeng and Tagliasacchi, Marco and Rybakov, Oleg and Ungureanu, Victor and Roblek, Dominik},
  booktitle={Proc. IEEE Int. Conf. Acoust., Speech, Signal Process. (ICASSP)},
  pages={691--695},
  year={2021}
}

@article{kim2019bandwidth,
  title={Bandwidth extension on raw audio via generative adversarial networks},
  author={Kim, Sung and Sathe, Visvesh},
  journal={arXiv preprint arXiv:1903.09027},
  year={2019}
}

@inproceedings{su2021bandwidth,
  title={Bandwidth extension is all you need},
  author={Su, Jiaqi and Wang, Yunyun and Finkelstein, Adam and Jin, Zeyu},
  booktitle={Proc. IEEE Int. Conf. Acoust., Speech, Signal Process. (ICASSP)},
  pages={696--700},
  year={2021}
}

@inproceedings{han2022nu,
  title={{NU-Wave 2}: A general neural audio upsampling model for various sampling rates},
  author={Han, Seungu and Lee, Junhyeok},
  year={2022},
  booktitle={Proc. Interspeech},
  pages={4401--4405},
  doi={10.21437/Interspeech.2022-45}
}

@inproceedings{li2015deep,
  title={A deep neural network approach to speech bandwidth expansion},
  author={Li, Kehuang and Lee, Chin-Hui},
  booktitle={Proc. IEEE Int. Conf. Acoust., Speech, Signal Process. (ICASSP)},
  pages={4395--4399},
  year={2015}
}

@inproceedings{abel2018simple,
  title={A simple cepstral domain {DNN} approach to artificial speech bandwidth extension},
  author={Abel, Johannes and Strake, Maximilian and Fingscheidt, Tim},
  booktitle={Proc. IEEE Int. Conf. Acoust., Speech, Signal Process. (ICASSP)},
  pages={5469--5473},
  year={2018}
}

@inproceedings{botinhao2006frequency,
  title={Frequency extension of telephone narrowband speech signal using neural networks},
  author={Botinhao, Cassia V and Carlos, Bruno S and Caloba, Luiz P and Petraglia, Mariane R},
  booktitle={Proc. Multiconf. Comput. Eng. Syst. Appl.},
  volume={2},
  pages={1576--1579},
  year={2006}
}

@inproceedings{shuai2023mdctgan,
  title={{mdctGAN}: Taming transformer-based {GAN} for speech super-resolution with modified {DCT} spectra},
  author={Shuai, Chenhao and Shi, Chaohua and Gan, Lu and Liu, Hongqing},
  year={2023},
  booktitle={Proc. Interspeech},
  pages={5112--5116},
  doi={10.21437/Interspeech.2023-113}
}

@inproceedings{lama,
  title={Resolution-robust large mask inpainting with {Fourier} convolutions},
  author={Suvorov, Roman and Logacheva, Elizaveta and Mashikhin, Anton and Remizova, Anastasia and Ashukha, Arsenii and Silvestrov, Aleksei and Kong, Naejin and Goka, Harshith and Park, Kiwoong and Lempitsky, Victor},
  booktitle={Proc. IEEE/CVF Winter Conf. Appl. Comput. Vis. (WACV)},
  pages={2149--2159},
  year={2022}
}

@inproceedings{gao2016snr,
  title={{SNR}-based progressive learning of deep neural network for speech enhancement},
  author={Gao, Tian and Du, Jun and Dai, Li-Rong and Lee, Chin-Hui},
  booktitle={Proc. Interspeech},
  pages={3713--3717},
  year={2016}
}

@article{hou2024snr,
  title={{SNR}-progressive model with harmonic compensation for low-{SNR} speech enhancement},
  author={Hou, Zhongshu and Lei, Tong and Hu, Qinwen and Cao, Zhanzhong and Lu, Jing},
  journal={IEEE Signal Process. Lett.},
  volume={32},
  pages={476--480},
  year={\rOne{2025}}
}

@inproceedings{jang2021univnet,
  title={{UnivNet}: A neural vocoder with multi-resolution spectrogram discriminators for high-fidelity waveform generation},
  author={Jang, Won and Lim, Dan and Yoon, Jaesam and Kim, Bongwan and Kim, Juntae},
  year={2021},
  booktitle={Proc. Interspeech},
  pages={2207--2211},
  doi={10.21437/Interspeech.2021-1016}
}

@article{dai2025sfnet,
  title={{SFNet}: A two-stage source-filter-based neural network for real-time speech bandwidth extension},
  author={Dai, Lingling and Ke, Yuxuan and Li, Andong and Li, Xiaodong and Zheng, Chengshi},
  journal={IEEE/ACM Trans. Audio, Speech, Lang. Process.},
  volume={34},
  pages={169--183},
  year={\rOne{2026}}
}

@inproceedings{rong2025ts,
  title={{TS-URGENet}: A three-stage universal robust and generalizable speech enhancement network},
  author={Rong, Xiaobin and Wang, Dahan and Hu, Qinwen and Wang, Yushi and Hu, Yuxiang and Lu, Jing},
  year={2025},
  booktitle={Proc. Interspeech},
  pages={863--867},
  doi={10.21437/Interspeech.2025-734}
}

@inproceedings{NISQA,
  author={Mittag, Gabriel and Naderi, Babak and Chehadi, Assmaa and M{\"{o}}ller, Sebastian},
  title={{NISQA}: A deep {CNN}-self-attention model for multidimensional speech quality prediction with crowdsourced datasets},
  booktitle={Proc. Interspeech},
  pages={2127--2131},
  year={2021},
  doi={10.21437/Interspeech.2021-299}
}

@article{PESQ,
  author={{ITU-T}},
  title={Perceptual evaluation of speech quality ({PESQ}): An objective method for end-to-end speech quality assessment of narrow-band telephone networks and speech codecs},
  journal={Rec. ITU-T P.862},
  year={2001}
}

@misc{vctk,
  title={{CSTR VCTK corpus}: English multi-speaker corpus for {CSTR} voice cloning toolkit (version 0.92)},
  author={Yamagishi, Junichi and Veaux, Christophe and MacDonald, Kirsten},
  year={2019},
  publisher={Univ. Edinburgh, The Centre for Speech Technology Research (CSTR)},
  howpublished={[Online]. Available: \url{https://datashare.ed.ac.uk/handle/10283/3443}}
}

@inproceedings{ears,
  title={{EARS}: An anechoic fullband speech dataset benchmarked for speech enhancement and dereverberation},
  author={Richter, Julius and Wu, Yi-Chiao and Krenn, Steven and Welker, Simon and Lay, Bunlong and Watanabe, Shinji and Richard, Alexander and Gerkmann, Timo},
  year={2024},
  booktitle={Proc. Interspeech},
  pages={4873--4877},
  doi={10.21437/Interspeech.2024-153}
}

@inproceedings{melgan,
  title={{MelGAN}: Generative adversarial networks for conditional waveform synthesis},
  author={Kumar, Kundan and Kumar, Rithesh and De Boissiere, Thibault and Gestin, Lucas and Teoh, Wei Zhen and Sotelo, Jose and De Brebisson, Alexandre and Bengio, Yoshua and Courville, Aaron C},
  booktitle={Adv. Neural Inf. Process. Syst. (NeurIPS)},
  volume={32},
  year={2019}
}

@inproceedings{mao2017least,
  title={Least squares generative adversarial networks},
  author={Mao, Xudong and Li, Qing and Xie, Haoran and Lau, Raymond YK and Wang, Zhen and Smolley, Stephen Paul},
  booktitle={Proc. IEEE Int. Conf. Comput. Vis. (ICCV)},
  pages={2794--2802},
  year={2017}
}

@inproceedings{lemercier2023analysing,
  title={Analysing diffusion-based generative approaches versus discriminative approaches for speech restoration},
  author={Lemercier, Jean-Marie and Richter, Julius and Welker, Simon and Gerkmann, Timo},
  booktitle={Proc. IEEE Int. Conf. Acoust., Speech, Signal Process. (ICASSP)},
  pages={1--5},
  year={2023},
  doi={10.1109/ICASSP49357.2023.10095258}
}

@article{lemercier2025review,
  title={Diffusion models for audio restoration: A review},
  author={Lemercier, Jean-Marie and Richter, Julius and Welker, Simon and Moliner, Eloi and V{\"a}lim{\"a}ki, Vesa and Gerkmann, Timo},
  journal={IEEE Signal Process. Mag.},
  volume={41},
  number={6},
  pages={72--84},
  year={2024},
  doi={10.1109/MSP.2024.3445871}
}

@inproceedings{shchekotov2022ffcse,
  title={{FFC-SE}: Fast {Fourier} convolution for speech enhancement},
  author={Shchekotov, Ivan and Andreev, Pavel and Ivanov, Oleg and Alanov, Aibek and Vetrov, Dmitry},
  booktitle={Proc. Interspeech},
  pages={1188--1192},
  year={2022},
  doi={10.21437/Interspeech.2022-603}
}

@article{kim2021multiresolution,
  title={A multi-resolution approach to {GAN}-based speech enhancement},
  author={Kim, Hyung Yong and Yoon, Ji Won and Cheon, Sung Jun and Kang, Woo Hyun and Kim, Nam Soo},
  journal={Applied Sciences},
  volume={11},
  number={2},
  pages={721},
  year={2021},
  doi={10.3390/app11020721}
}

@inproceedings{visqol,
  title={{ViSQOL v3}: An open source production ready objective speech and audio metric},
  author={Chinen, Michael and Lim, Felicia S. C. and Skoglund, Jan and Gureev, Nikita and O'Gorman, Feargus and Hines, Andrew},
  booktitle={Proc. Int. Conf. Quality Multimedia Experience (QoMEX)},
  pages={1--6},
  year={2020},
  doi={10.1109/QoMEX48832.2020.9123150}
}

@article{itut_p8622_2007,
  author={{ITU-T}},
  title={{Wideband extension to \rOne{Recommendation P.862} for the assessment of wideband telephone networks and speech codecs}},
  journal={Rec. ITU-T P.862.2},
  year={2007}
}

\end{document}